\documentclass[a4paper,10pt,aps,prd,superscriptaddress,floatfix]{revtex4-1}
\usepackage{graphicx}
\usepackage{amsmath}
\usepackage{mathptmx}
\usepackage{multirow}
\usepackage{subfigure}
\usepackage{rotating}
\usepackage{longtable}
\usepackage{array}

\def\e{\epsilon}

\begin{document}

\title{Digital watermarking : An approach based on Hilbert transform}

\author{Rashmi Agarwal}
\email{t.rashmiagarwal@gmail.com}
\affiliation{Department of Computer Science, Mohan Lal Sukhadia University, Udaipur, India-313 039.}
\author{R. Krishnan}
\email{rkrishnan@imsc.res.in}
\affiliation{The Institute of Mathematical Sciences, CIT Campus, Taramani, Chennai, India-600 113.}
\author{M. S. Santhanam}
\email{santh@iiserpune.ac.in}
\affiliation{Indian Institute of Science Education and Research, Pune, India-410 021.}
\author{K. Srinivas}
\email{srini@imsc.res.in}
\affiliation{The Institute of Mathematical Sciences, CIT Campus, Taramani, Chennai, India-600 113.}
\author{K. Venugopalan}
\email{venumlsu@gmail.com}
\affiliation{Department of Physics, Mohan Lal Sukhadia University, Udaipur, India-313 039.}
\begin{abstract}
\begin{center}
{\small ABSTRACT}
\end{center}
Most of the well known algorithms for watermarking of digital images
involve transformation of the image data to Fourier or singular
vector space. In this paper, we introduce watermarking in Hilbert
transform domain for digital media. Generally, if the image is a
matrix of order $m$ by $n$, then the transformed space is also an image
of the same order. However, with Hilbert transforms, the transformed
space is of order $2m$ by $2n$. This allows for more latitude in
storing the watermark in the host image. Based on this idea, we propose
an algorithm for embedding and extracting watermark in a host image
and analytically obtain a parameter related to this procedure.
Using extensive simulations, we show that the algorithm performs
well even if the host image is corrupted by various attacks.
\end{abstract}
\maketitle

\section{Introduction}

A large amount of accessible information today is available in one
or the other multimedia formats. While this serves the purpose of
easier dissemination of data, this also makes it vulnerable for
misappropriation and misuse. Hence, it is all the more important to
protect intellectual property rights of the content available in
various digital formats. Digital watermarking \cite{Langelaar,Cox-1} is a popular method by
which the owners of the data, in any multimedia format, can embed
their logo, trademark or some proprietary information \cite{Zeng-2,Petitcolas} in a way that
can either be visible or invisible to a general user. This
information can be later retrieved for verification purposes or in
case of conflicting claims on the ownership of data \cite{Craver,Piva-2,Zeng-1}.

Thus, when applied to the case of digital images, digital watermarking
technique consists of (i) an algorithm to embed a watermark image on
a host image and (ii) an algorithm to retrieve the embedded
watermark with least distortion. Ideally, we would expect that the
algorithms be robust against any manipulation of the original data
and it should also be designed to render any illegal retrieval of
watermark a futile exercise. The field of digital watermarking has
been the focus of research attention for more than a decade now.
This is partly due to the proliferation of multimedia formats as
well as new tools, both commercial and open source, to manipulate
them. In this paper, we focus on the invisible watermarking of
digital images using the Hilbert transform technique.

Watermarking techniques for digital images can be broadly classified
into two categories, namely, the spatial domain techniques and
transform domain techniques depending on which domain the watermark
is embedded. Typically, in spatial domain techniques the watermark
is embedded in those part of the data that do not distort the host
image in any significant way. For instance, some of the well-known
spatial domain techniques are least significant substitution
\cite{Cox-2,Podilchuk} and the correlation based
approach \cite{Kallel-2,Raval}. In least significant substitution
technique, the watermark is embedded by replacing the least
significant bits of the image data with the bits of the watermark
data. There are many variants of this technique. In correlation
based approach the watermark is converted to a pseudo-random noise
(PN) sequence which is then weighted and added to the host image
with a gain factor. For detection, the watermarked image is
correlated with the watermark image. In the transform domain
techniques, the watermark is embedded in those parts of the
transformed host image which do not distort the image significantly.

One of the earliest transform domain techniques is the one based on
discrete cosine transform (DCT) \cite{Barni-1,Chu,Hernandez,Dickinson,Piva-1}.
In DCT, the image is
decomposed in terms of various frequency bands and watermarks are
embedded in the middle frequency bands which are not significant for
the host image. Further, image transformations do not affect the
watermark placed in those bands. DCT based methods are generally
robust, particularly against JPEG and MPEG compression. The
techniques based on wavelet decomposition are similar in spirit to
DCT with the additional feature that the multi-resolution character
of the wavelets allows graded information to be stored at various
resolutions. For instance, in \cite{Kundur} wavelet coefficients of
the image and the watermark at different levels of resolution are
added together within the constraint of the so-called human-visual
model. A review of wavelets based techniques is available in
\cite{Barni-2,Wang,Karras,Taweel,Kundur-2}. There is yet another
method of digital watermarking based on singular value decomposition
(SVD) techniques
\cite{Chang-1,Chang-2,Liu,Gorodetski,Chandra,Sun,Yongdong,Agarwal,Shieh,Andrews}.
In contrast to DCT and wavelets based techniques, the advantage of
singular value decomposition based methods is that they provide a
transform space that is tailor made for the given image data matrix.
Both in the DCT and wavelets, the basis for the transform space is a
fixed set of functions. In the SVD, it must be calculated from the
given data and the singular vectors so calculated form an optimal
basis for the image matrix in the least square sense. It is worth mentioning that
some authors have resorted to hybrid techniques i.e., algorithms based simultaneously
on different domains to improve the watermarking results \cite{Po,Ganic-1,Ganic-2,Gaurav}.

In this present work, we propose a new scheme for watermarking
digital images using the Hilbert transform. The analytic signal
$\hat{s}(t)$ associated to a signal $s(t)$ is,
\begin{equation}
\hat{s}(t) = s(t) + i s_H(t)
\end{equation}
where $s_H(t)$ is the Hilbert transform of $s(t)$. Clearly,
$\hat{s}(t)$ can be written in phase-amplitude form as $A(t)  e^{i
\theta(t)}$. If the signal changes sufficiently slowly, then the
phase of the analytic signal is negligible. Typically, most images
have slowly varying pixel values except at the edges.
In such a scenario, we can expect the phase to be
negligible most of the time and the matrix of phase values will be
sparse. Hence a good amount of information can be embedded in the
phase of the analytic signal associated to the image. Since only the
phase of the analytic signal is proposed to be used for embedding
the watermark, it is likely to be highly imperceptible for visual
perception. This is one of the key requirements of ideal
watermarking algorithms. In addition, this also provides a large
space for embedding the watermark which is useful for creating
redundant watermark distributed throughout the Hilbert transformed
space. This makes the algorithm more robust against attacks. This is
the main idea underlying the present scheme and to the best of our
knowledge the Hilbert transform has not been used for watermarking
purposes before.

Further, our proposal is a non-blind scheme, which implies that to
recover the watermark from the watermarked image, we require both
the original as well as the watermark image. This is not necessarily
a restrictive requirement as there are many situations in which this
scenario is valid, such as in ownership litigations. In many such
cases, the owner keeps a copy of both the original host image and
the watermark. In addition, non-blind scheme also makes detection of
watermark in an arbitrary image difficult if the watermark image is
not available.

The rest of the paper is organized as follows.
In the next section, we introduce the proposed method of digital watermarking.
In Sec.III, we provide the formula to optimize the scaling factor. Section IV
describes the results of  numerical simulation, while Sec. V describes
robustness of the algorithm. Section VI gives a summary and concluding remarks.

\section{Watermarking using the Hilbert transform}

Given any arbitrary signal $s(t)$, as a function of time $t$, we can construct an analytic
signal of the form
\begin{equation}
\hat{s}(t) = s(t) + i s_H(t) = A(t) e^{i \theta(t)}
\end{equation}
where $s_H(t)$ is the Hilbert transform of $s(t)$. It is defined in terms of
the Cauchy principal value (P.V) of the integral
\begin{equation}
s_H(t) = \frac{1}{\pi} \mbox{P.V} \int_{-\infty}^{\infty} \frac{s(\tau)}{t-\tau} ~d\tau
\end{equation}
provided this integral exists. In general, Hilbert transform has a
wide range of applications, in particular, in the area of signal
processing \cite{Stefan}. In signal processing, the Hilbert transform $\bf{z_H}$
of a discrete signal $\bf{z}$ is defined as the output of a linear
filter with the frequency response $H(\omega)$ given by
\begin{equation}
 H(\omega) = \left\{
\begin{array}{l l}
  i, & \quad -\pi <\omega <0\\
  -i, & \quad 0<\omega < \pi\\
  0, & \quad \omega=-\pi, 0, \pi\\
\end{array} \right.
\end{equation}
In the context of this work, let $\mathbf{z}$ denote a signal vector
of size $n \times 1$ as a function of position (in arbitrary units).
For instance, $\mathbf{z}$ would represent one column of pixel
values of an image matrix $\mathbf{Z}$ of order $n \times m$. Then,
the analytic signal $\hat{z}$ associated with $\mathbf{z}$ is $\bf{z
+ i z_H}$, a complex valued signal with phase and amplitude. We
intend to embed the watermark image in the phase component. We
denote the amplitude and phase of this analytic signal by $n \times
1$ vectors $\mathbf{r_z}$ and $\mathbf{\theta_z}$ respectively. The
inverse Hilbert transform consists of taking the real part
$\mathbf{r_z}$ of this analytic signal. Hence we have,
\begin{equation}\label{hilbert}
\mathbf{z = r_z * \cos \theta_z},
\end{equation}
where the $\mathbf{*}$ stands for element-wise multiplication and is not to
be confused with usual matrix multiplication. Further, for any vector $\mathbf{\theta_z}$,
the operation $\cos \mathbf{\theta_z}$ represents the cosine of every element of the vector.

Notice that the Hilbert transforms in two dimensions is not uniquely
defined. Hence, in the watermarking algorithm presented below, we
will treat the image matrix $\mathbf{Z}$ of size $n \times m$ as $n$
vectors and apply Hilbert transforms to each of the $n$ vectors.
From a computational point of view, for the Hilbert transform of any
image $\mathbf{Z}$, we are required to compute $n$ independent
Hilbert transforms each with $m$ elements. However, for ease of
notation, we present the algorithm using scalar elements instead of
in vector notation. Now we will apply this to define Hilbert
transform of an image matrix $\mathbf{Z}$ of size $n \times m$ with
elements $Z_{ij}, i=1,2,3,\ldots n, j=1,2,3,\ldots m$. Hence
Eq.\eqref{hilbert} will still continue to hold in the form
\begin{equation} \label{hilbertmat}
Z_{ij} = a_{ij}  \cos\Theta_{ij}, \quad i=1,2,3, \ldots n, \quad
j=1,2,3,\ldots m.
\end{equation}
In this, $a_{ij}$ are the amplitudes and $\Theta_{ij}$ are the phases
obtained from vector-wise Hilbert transform applied on $\mathbf{Z}$.

\subsection{Algorithm for embedding the watermark}

In this section we describe the algorithm to embed a watermark image into
another gray scale image of the same size using the Hilbert transform. Let
the host image be represented by $\mathbf{Z}$ with elements
$Z_{i j}$ and $\mathbf{W}$ denote
the matrix of the watermark image with elements $W_{ij}$.

First, we perform the Hilbert transform of the original image $\mathbf{Z}$
and obtain the relation
\begin{equation}
Z_{ij} = a_{ij}  \cos\Theta_{ij}. \;\;\;\;\; \label{hilbert_Z}
\end{equation}
Next, we do the same for the watermark image $\mathbf{W}$ to get
\begin{equation}
W_{ij} = b_{ij}  \cos\Phi_{ij}, \label{hilbert_W}
\end{equation}
where $b_{ij}$ and $\Phi_{ij}$ are the phase and amplitude.

Now, we add the scaled amplitude $b_{ij}$ of the watermark to the
phase $\Theta_{ij}$ of the original image to get
\begin{equation}\label{new_phase}
\theta_{ij} = \Theta_{ij} + \lambda b_{ij},
\end{equation}
where $\lambda$ is the scaling factor. For typical images, the order
of magnitude of $b_{i j}$'s is much larger than $\Theta{i j}$'s for
all $i,j$ and hence $\lambda$ will have to be much smaller than
unity in order to compensate for this difference in order of
magnitudes.

Finally, we get the watermarked image $\mathbf{Z^w}$ as
\begin{equation}\label{embedding}
Z^w_{ij} = a_{ij} \cos\theta_{ij}.
\end{equation}
Thus equations \eqref{hilbert_Z}-\eqref{embedding} constitute the algorithm for
watermarking using the Hilbert transform applied column-wise to an image matrix.
We remark that this algorithm is motivated by the fact that most of the
information about the image is encapsulated within the amplitude.
The phases $\Theta_{ij}$, for all $i$ and $j$, contain very little information and can be thought
of as a sparse matrix. Hence, we can store most of the information about the
watermark without causing too much distortion in the original image by following
the above strategy.

\subsection{Algorithm for extracting the watermark}

Given the watermarked image $\mathbf{Z^w}$, we can extract a
(possibly corrupted) watermark if we have access to $\Theta_{ij},
\Phi_{ij}, a_{ij}$, for all $i$ and $j$, and the value of $\lambda$.
This information is most easily available if one has access to the
original image as well as the watermark image. As pointed out in a previous
work of the first and the third author \cite{Agarwal}, this is not a
particularly restrictive assumption.

The extraction algorithm is just the reversal of the embedding
algorithm given in the previous subsection. Starting from Eq.
\eqref{embedding} we divide both sides by $a_{ij}$ and use the
inverse cosine function to recover $\theta_{ij}$. By substituting
for $\theta_{ij}$ using Eq. \eqref{new_phase} we get for the
amplitude of the watermarked image
\begin{equation}\label{extracted_amplitude}
{\widetilde{b}}_{ij} = \frac{\cos^{-1}\left( Z^w_{ij} /a_{ij} \right) -
\Theta_{ij}}{\lambda}.
\end{equation}
Finally using Eq. \eqref{hilbert_W}, the extracted watermark image
can be constructed as follows,
\begin{equation}\label{extraction}
\widetilde{W}_{ij} = \widetilde{b}_{ij} \cos (\Phi_{ij}).
\end{equation}
Thus, Eq. \eqref{extraction}, along with Eq. \eqref{extracted_amplitude} constitute
the watermark extraction algorithm.

\section{Optimization of the scaling factor}

The choice of the value of the scaling factor $\lambda$ plays an important role in
our watermarking algorithm. If $\lambda$ is chosen too small, then the quality of
embedding is good but that of the extracted watermark is poor. On the other hand,
if $\lambda$ is too large, then extraction works well but embedding suffers. Hence,
$\lambda$ needs to be chosen optimally to achieve a balance between these extremes.
In general, this problem is a non-linear optimization problem and we shall describe
an iterative method to produce a solution to it.

We shall use the mean square error (MSE) as a measure of the quality of embedding
or extraction, and attempt to minimize the sum of the MSE from the embedding and
extraction steps. First define the function $f(\lambda)$ as
\begin{equation}
f(\lambda) = \sum_{j=1}^{m} \sum_{i=1}^{n} \Big[(Z^w_{ij} - Z_{ij})^2
+ (\widetilde{W}_{ij}-W_{ij})^2 \Big].
\end{equation}
With this notation, we need to find the value of $\lambda$ that
minimizes $f(\lambda)$. Note that both $\mathbf{Z^w}$ and
$\mathbf{\widetilde{W}}$ depend on $\lambda$. We also note that, in
practice, converting $\mathbf{Z^w}$ to an image introduces a
truncation error $\epsilon$, which, though negligible in the
embedding step, affects the extraction step significantly.

Now, we use Eqs. \eqref{new_phase} and \eqref{embedding} adjusted for the truncation error $\epsilon$ to get the following
\begin{align} \nonumber
\sum_{j=1}^{m} \sum_{i=1}^{n} (Z^w_{ij} - Z_{ij})^2  &=
\sum_{j=1}^{m} \sum_{i=1}^{n} \big[a_{ij}\cos(\Theta_{ij} + \lambda
b_{ij}) \\ \label{mse1}
&+\epsilon- a_{ij} \cos\Theta_{ij}\big]^2 \\
&\approx \sum_{j=1}^{m} \sum_{i=1}^{n} (-\lambda a_{ij} b_{ij} \sin\Theta_{ij} + \ldots)^2, \nonumber
\end{align}
where the terms involving $\epsilon$ and higher powers of $\lambda$ are ignored.

Similarly we use Eqs. \eqref{extracted_amplitude} and \eqref{extraction} adjusted for $\epsilon$ to get
\begin{align}\nonumber
& \sum_{j=1}^{m} \sum_{i=1}^{n} (\widetilde{W}_{ij}-W_{ij})^2 \\ \nonumber
&= \sum_{j=1}^{m} \sum_{i=1}^{n} \bigg[\frac{\cos^{-1}
\big[\cos(\Theta_{ij} + \lambda b_{ij})+\epsilon/a_{ij} \big]
-\Theta_{ij}}{\lambda} \\ \nonumber
& \times \cos \Phi_{ij} - b_{ij} \cos \Phi_{ij} \Big]^2 \\ \label{mse2}
&=\sum_{j=1}^{m} \sum_{i=1}^{n} \left(\frac{\alpha(\lambda) \cos
\Phi_{ij} }{\lambda} \right)^2 ,
\end{align}
where $\alpha(\lambda)$ is given by
\begin{equation}\label{alpha}
\alpha(\lambda) = \cos^{-1} (\cos(\Theta_{ij} + \lambda b_{ij})+\epsilon/a_{ij})
-(\Theta_{ij} + \lambda b_{ij}).
\end{equation}
With this preparation we return to the minimization of $f(\lambda)$.
Our approach is to start with any initial value $\lambda=\lambda_0$
and produce successive values $\lambda_1, \lambda_2, \ldots$, each
depending on the previous value, which converge to the desired
minimum of $f(\lambda)$.

The method to produce $\lambda_{\ell+1}$ from $\lambda_\ell$ at the $\ell$-th step is as follows.
First we use Eqs. \eqref{mse1} - \eqref{alpha} to approximate $f(\lambda)$ by $f_\ell(\lambda)$ as,
\begin{equation}
f_\ell(\lambda) = \sum_{j=1}^{m} \sum_{i=1}^{n} \Big[(a_{ij} b_{ij} \sin
\Theta_{ij})^2 \lambda^2
+ (\alpha(\lambda_\ell) \cos \Phi_{ij})^2 \lambda^{-2}\Big].
\end{equation}
Note that $f_\ell(\lambda)$ depend on $\lambda_\ell$. Next, we choose $\lambda_{\ell+1}$ to be the value at which $f_\ell(\lambda)$ attains its minimum.
After a straightforward computation, we have the following,
\begin{equation}\label{optimization}
 \lambda_{\ell+1} = \left(\frac{\sum (\alpha(\lambda_\ell) \cos \Phi_{ij} )^2}
{\sum (a_{ij} b_{ij} \sin \Theta_{ij})^2}\right)^{1/4}.
\end{equation}
Thus Eq. \eqref{optimization} can be iterated for $\ell=0,1,\ldots$
starting from any initial value $\lambda_0$ till the desired level
of convergence is achieved. In practice, this algorithm seems to
have fast convergence. For
example, for our test images it converges within $\ell=2$ steps to
$\lambda=0.0018$ (upto 4 decimal places) for a wide range of initial
values.

\section{Numerical simulation}

\begin{figure}
\begin{center}
\subfigure[]{\label{dollar512} \includegraphics[width=1.3in,height=1.3in]{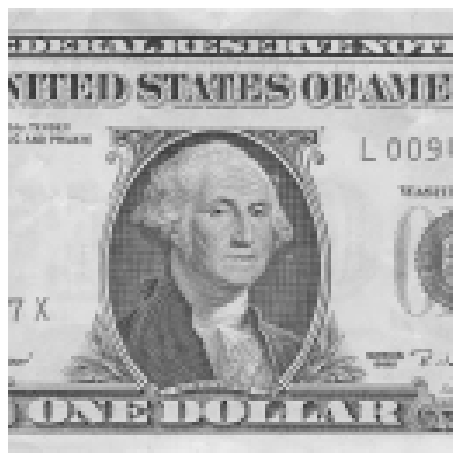}}
\hspace*{4mm}
\subfigure[]{\label{peppers512} \includegraphics[width=1.3in,height=1.3in]{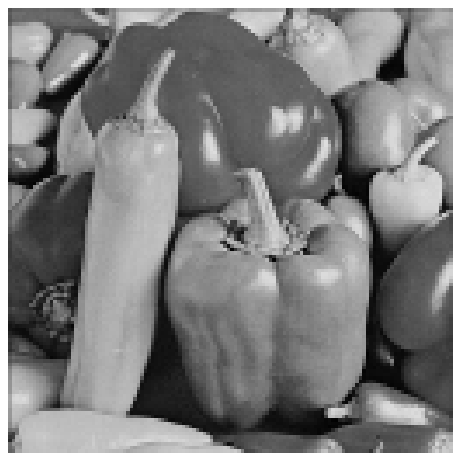}}
\end{center}
\caption{\subref{dollar512} Host image and \subref{peppers512}
watermark image of size 512 $\times$ 512.}
\label{demo1}
\end{figure}

\begin{figure}
\begin{center}
\subfigure[]{\label{dollar512_EMB}\includegraphics[width=1.3in,height=1.3in]{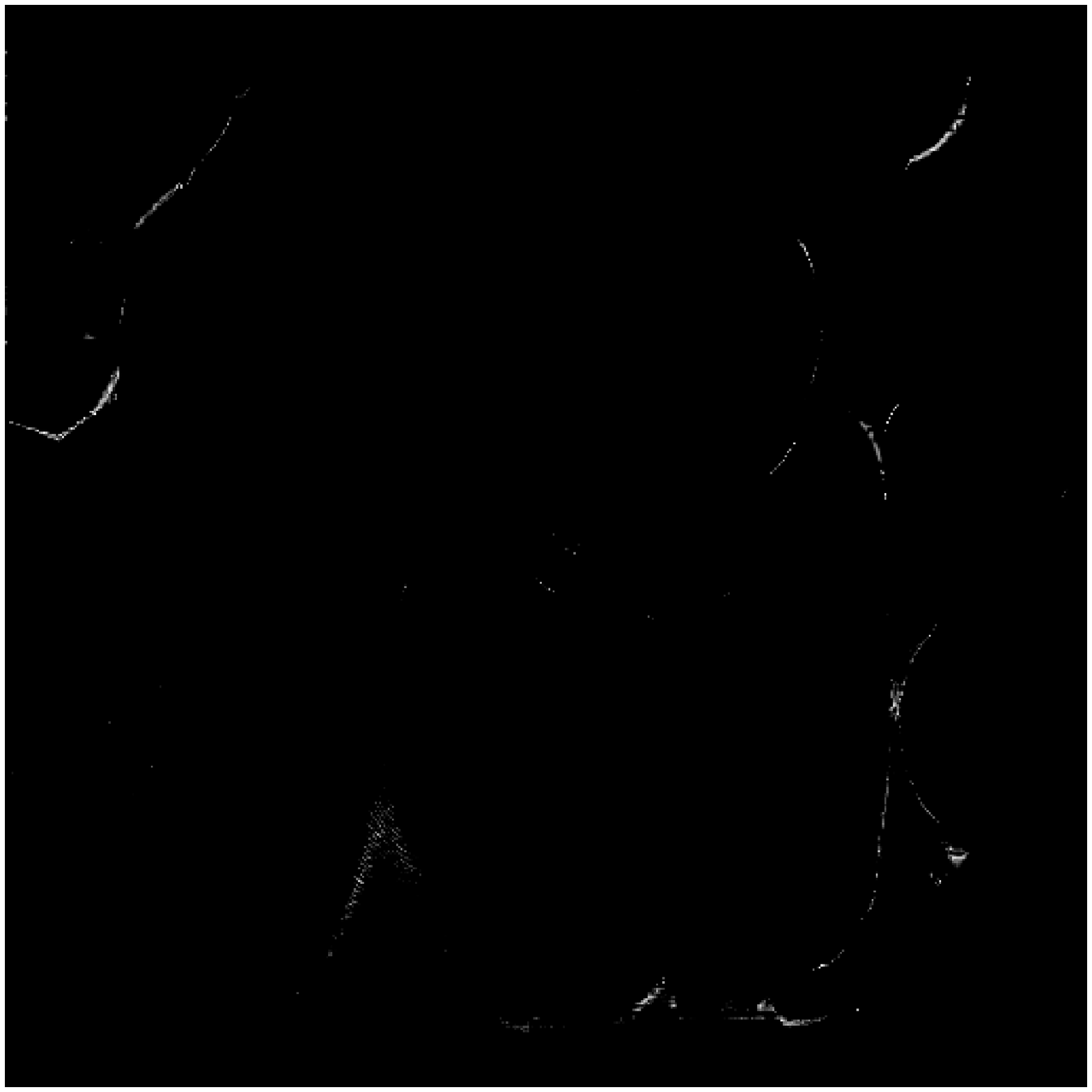}}
\hspace*{4mm}
\subfigure[]{\label{peppers512_EXT}\includegraphics[width=1.3in,height=1.3in]{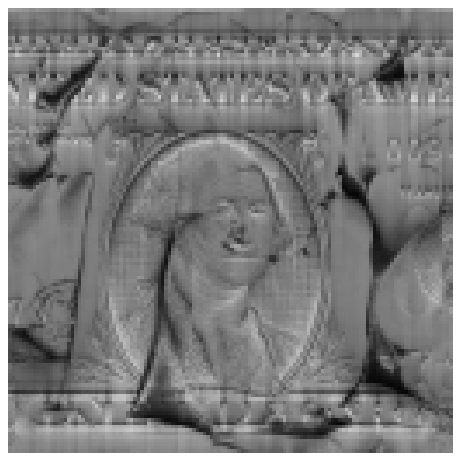}}
\\
\subfigure[]{\label{d_EMB_0.01}\includegraphics[width=1.3in,height=1.3in]{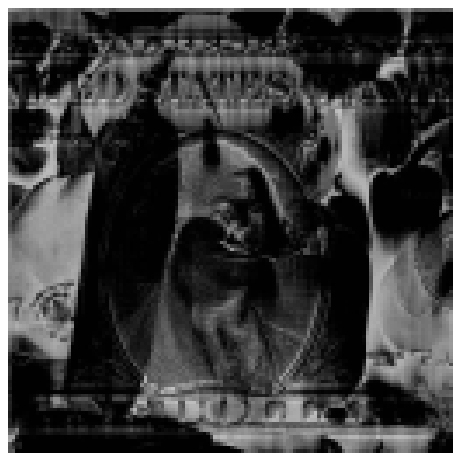}}
\hspace*{4mm}
\subfigure[]{\label{p_EXT_0.01}\includegraphics[width=1.3in,height=1.3in]{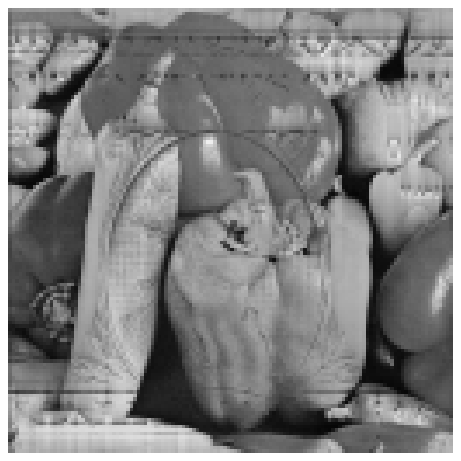}}
\\
\subfigure[]{\label{d_EMB_0.0018}\includegraphics[width=1.3in,height=1.3in]{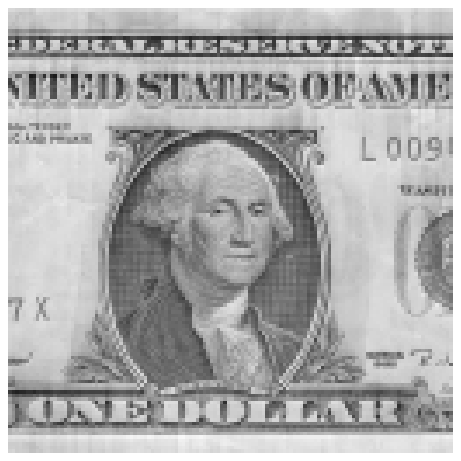}}
\hspace*{4mm}
\subfigure[]{\label{p_EXT_0.0018}\includegraphics[width=1.3in,height=1.3in]{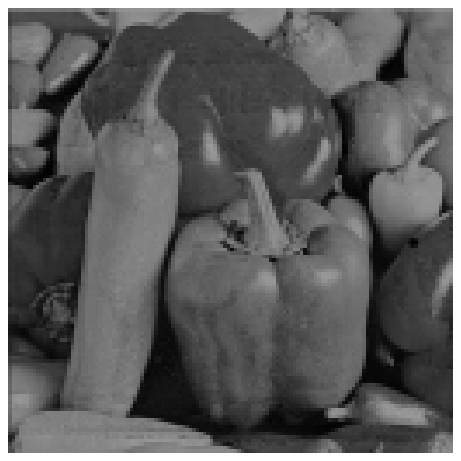}}
\\
\subfigure[]{\label{d_EMB_0.001}\includegraphics[width=1.3in,height=1.3in]{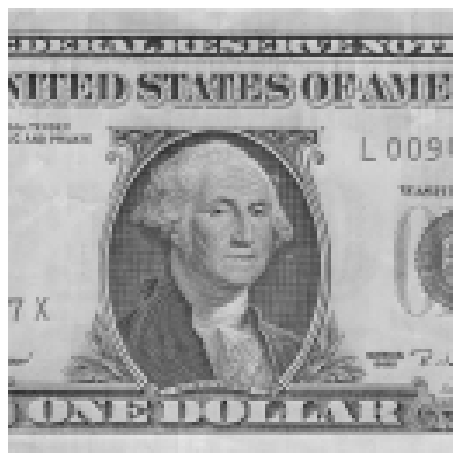}}
\hspace*{4mm}
\subfigure[]{\label{p_EXT_0.001}\includegraphics[width=1.3in,height=1.3in]{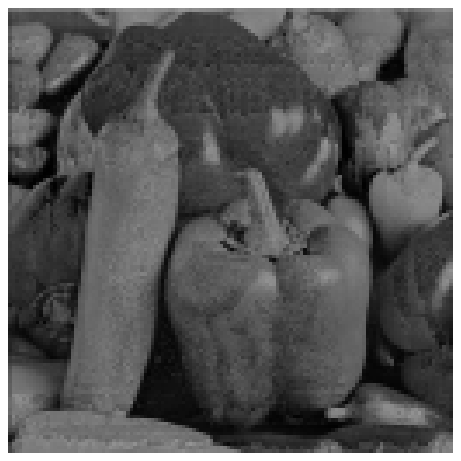}}
\\
\subfigure[]{\label{d_EMB_0.0001}\includegraphics[width=1.3in,height=1.3in]{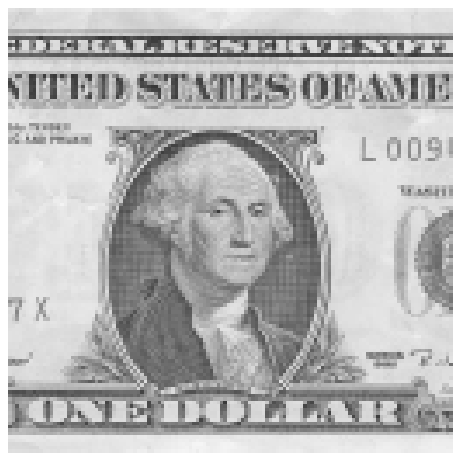}}
\hspace*{4mm}
\subfigure[]{\label{p_EXT_0.0001}\includegraphics[width=1.3in,height=1.3in]{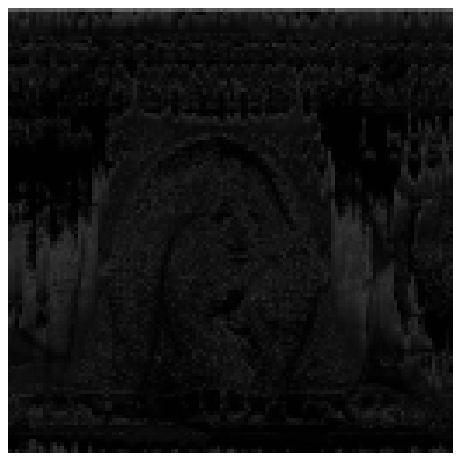}}
\caption{This panel shows the watermarked image on the left panel
and the extracted watermark on the right panel for
\subref{dollar512_EMB}, \subref{peppers512_EXT} $\lambda = 0.1$,
\subref{d_EMB_0.01}, \subref{p_EXT_0.01} $\lambda = 0.01$,
\subref{d_EMB_0.0018}, \subref{p_EXT_0.0018} $\lambda = 0.0018$,
\subref{d_EMB_0.001}, \subref{p_EXT_0.001} $\lambda = 0.001$ and
\subref{d_EMB_0.0001}, \subref{p_EXT_0.0001} $\lambda = 0.0001$.}
\label{demo2}
\end{center}
\end{figure}

In this section we will apply our embedding and extraction algorithm
to the host image shown in Fig. \ref{dollar512} and the watermark
image in Fig. \ref{peppers512}. Both these images have dimensions
$512 \times 512$. As pointed out earlier, the quality of the
watermarked image improves and that of the extracted image
deteriorates as the scaling parameter $\lambda \to 0$. This is borne
out by the simulation results and can be clearly seen from the
watermarked images shown in Figs. \ref{dollar512_EMB},
\ref{d_EMB_0.01}, \ref{d_EMB_0.0018}, \ref{d_EMB_0.001}, and
\ref{d_EMB_0.0001} for $\lambda=0.1, 0.01, 0.0018, 0.001, 0.0001$
respectively. The corresponding extracted watermark images are shown
in Figs. \ref{peppers512_EXT}, \ref{p_EXT_0.01}, \ref{p_EXT_0.0018},
\ref{p_EXT_0.001}, and \ref{p_EXT_0.0001}. The simulations clearly demonstrate that the Hilbert transform
based algorithm proposed in Eq. \eqref{hilbert_Z}-\eqref{embedding}
and Eq. \eqref{extracted_amplitude},\eqref{extraction} produces
results whose visual quality is good and acceptable as a practical
tool for watermarking digital images.

\begin{figure}
\begin{center}
\begin{turn}{-90}
\includegraphics[width=2.0in,height=1.6in]{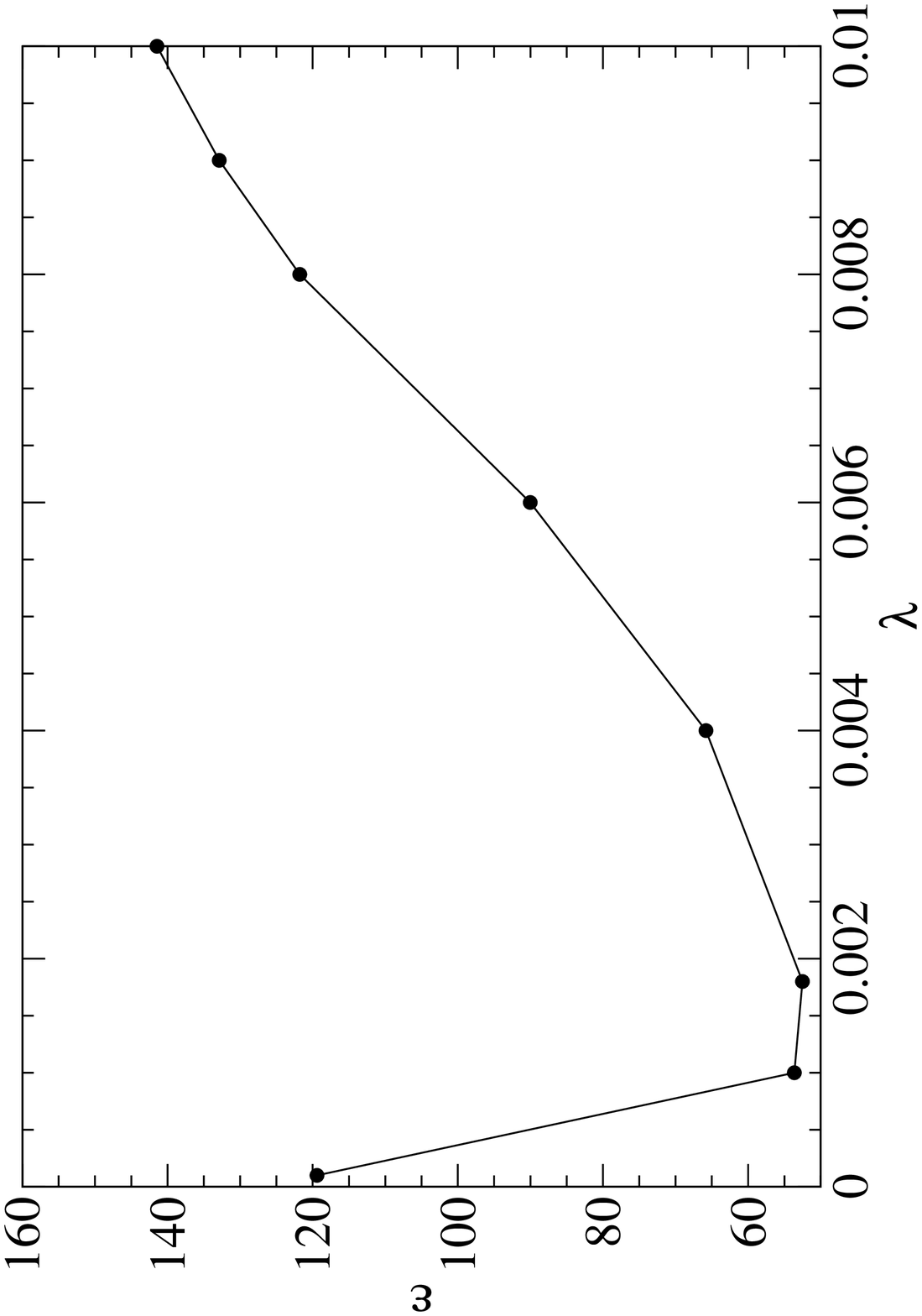}
\end{turn}
\end{center}
\caption{Mean square error versus $\lambda$ plot.} \label{rmse-plot}
\end{figure}

These results are quantified by means of
peak signal to noise ratio (PSNR) and the root-mean-square error
(RMSE) of the corresponding images.
PSNR is used to evaluate the perceptual distortion of the proposed
scheme. PSNR and RMSE is computed for the image difference matrix
$\Delta_{ij} = I_{ij} - \widetilde{I}_{ij}$ for several values of $\lambda$.
In our context, there are two possibilities, (a) $I_{ij}$ are the elements of host image and
$\widetilde{I}_{ij}$ are the elements of watermarked image, and (b) $I_{ij}$ are the
elements of watermark image and $\widetilde{I}_{ij}$ are the elements of extracted
watermark.
Then RMSE is defined as
\begin{equation} \label{rmse-eq}
\varepsilon=\sqrt{\frac{1}{mn}\sum_{i=1}^m \sum_{j=1}^n
\Delta_{i,j}^2} \,.
\end{equation}
We also define the PSNR, measured in decibels, as
\begin{equation} \label{psnr-eq}
p = 10 \log_{10}(\mbox{max}~ z_{i,j}^2/\varepsilon^2),
\end{equation}
where $\mbox{max}~ z_{i,j}$ represents the maximum value of a matrix
whose elements are $z_{i,j}$. Figure \ref{rmse-plot} depicts the combined RMSE versus lambda graph of the extracted and watermarked images.
Notice that $\lambda=0.0018$ corresponds to a minima as predicted by the analysis in the previous section.


\section{Robustness of the algorithm}


An important property of the watermarking algorithms is that they
should be robust against various possible attacks. In this section,
we will first embed the watermark in the host. The watermarked image
will then be subjected to some attacks. After the attack has been
carried out, the extraction algorithm is applied. We will
subject our algorithm to the following major attacks (i) robustness
against Additive Noise (ii) Cropping (iii) Gaussian Noise (iv) JPEG
and JPEG 2000 compression (v) Median Filter (vi) Rotation (vii) Gamma Correction
(viii) Intensity Adjustment (ix) Gaussian Blur (x) Contrast Enhancement
(xii) Dilation (xiii) Scaling.
The full list of attacks and their result is given in Table. (\ref{Hilbert-tab}).
Before we proceed, at the very outset let us mention that all the
images are taken to be of $512 \times 512$ and the scaling factor is
chosen to be $0.0018$ (optimal value). In particular, all the attacks
are done on Fig. \ref{d_EMB_0.0018}.

\subsection{Additive Noise}

\begin{figure}
\begin{center}
\subfigure[]{\label{d-noise}\includegraphics[width=1.3in,height=1.3in]{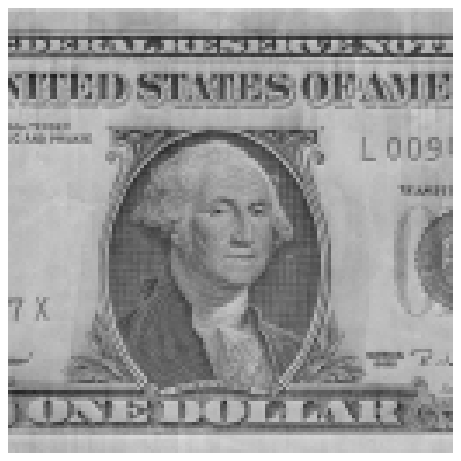}}
\hspace*{4mm}
\subfigure[]{\label{p-noise}\includegraphics[width=1.3in,height=1.3in]{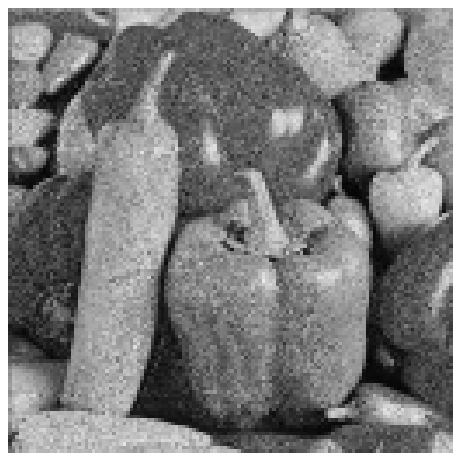}}
\end{center}
\caption{\subref{d-noise} Noise added watermarked image and
\subref{p-noise} Extracted watermark.} \label{noise}
\end{figure}

Our first attack will be the addition of uniformly distributed noise to Fig.
\ref{d_EMB_0.0018} resulting in the corrupted image shown in Fig.
\ref{d-noise}. Figure \ref{p-noise} shows the result of extracting
the watermark. As can be noticed visually and also from the $p$ and
$\e$ values, the quality of the extracted watermark is quite good.

\subsection{Cropping}

\begin{figure}
\begin{center}
\subfigure[]{\label{d-crop}\includegraphics[width=1.3in,height=1.3in]{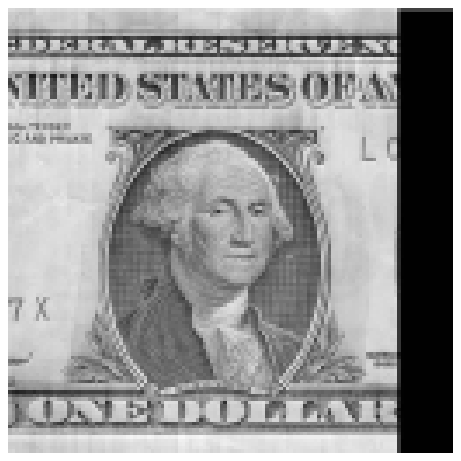}}
\hspace*{4mm}
\subfigure[]{\label{p-crop}\includegraphics[width=1.3in,height=1.3in]{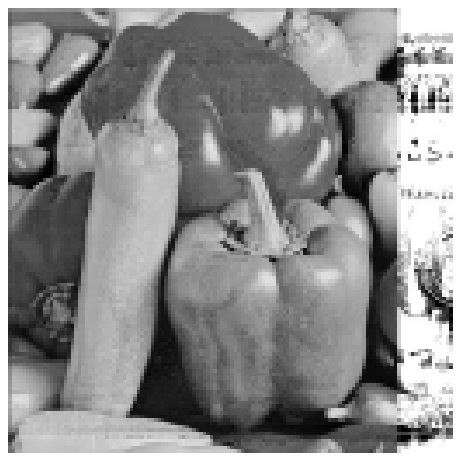}}
\caption{\subref{d-crop} Cropped watermarked image and
\subref{p-crop} Extracted watermark.} \label{crop}
\end{center}
\end{figure}

Cropping is the process of removing a portion of the image. In the present
case we will remove a portion of the watermarked image and check if our
extraction algorithm is able to extract the watermark. Figures \ref{d-crop} and
\ref{p-crop} show the cropped watermarked images and the extracted watermark
respectively. Though there is
some residual effect of the host in the extracted image, it is worth
noting that it appears only in the cropped portion of the image. The
extracted image is unaffected by this residual effect of the host.
We would also like to mention that we have carried out cropping for
various percentages ($10\%$, $20\%$, $30\%$, $40\%$, $50\%$) of Fig.
\ref{d_EMB_0.0018} and the extracted image is good.

\subsection{Gaussian Noise}

\begin{figure}
\begin{center}
\subfigure[]{\label{d-gaussian}\includegraphics[width=1.3in,height=1.3in]{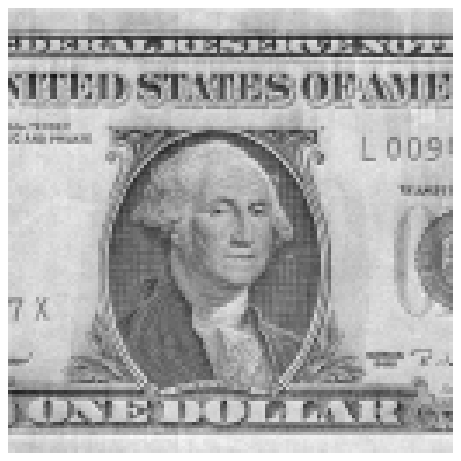}}
\hspace*{4mm}
\subfigure[]{\label{p-gaussian}\includegraphics[width=1.3in,height=1.3in]{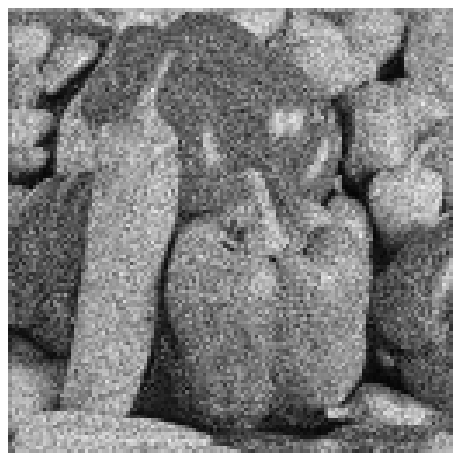}}
\caption{\subref{d-gaussian} Gaussian noise watermarked image and
\subref{p-gaussian} Extracted watermark.} \label{g-noise}
\end{center}
\end{figure}

When Gaussian distributed noise is added to the watermarked image we obtain Fig.
\ref{d-gaussian}. The extracted image from this is shown in Fig.
\ref{p-gaussian}. Once again our extraction algorithm performs well
and yields the watermark.

\subsection{JPEG and JPEG 2000 Compression}

\begin{figure}
\begin{center}
\subfigure[]{\label{d-20-jpeg}\includegraphics[width=1.3in,height=1.3in]{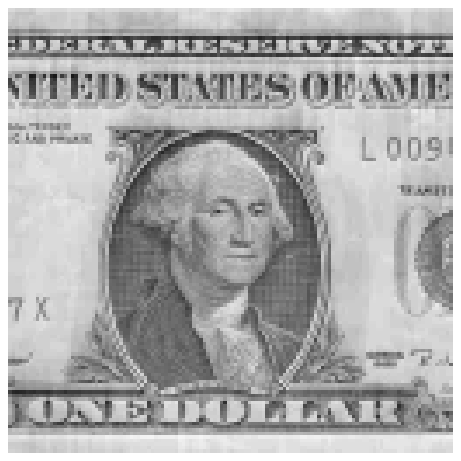}}
\hspace*{4mm}
\subfigure[]{\label{p-20-jpeg}\includegraphics[width=1.3in,height=1.3in]{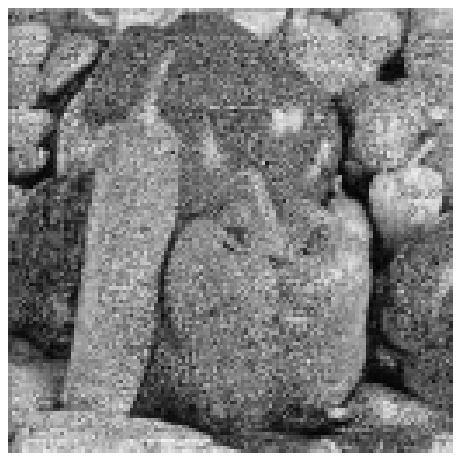}}
\\
\subfigure[]{\label{d-40-jpeg}\includegraphics[width=1.3in,height=1.3in]{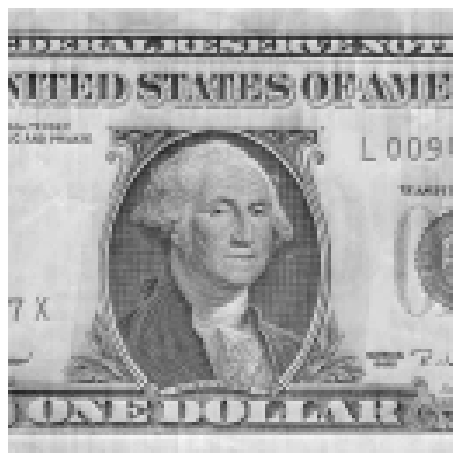}}
\hspace*{4mm}
\subfigure[]{\label{p-40-jpeg}\includegraphics[width=1.3in,height=1.3in]{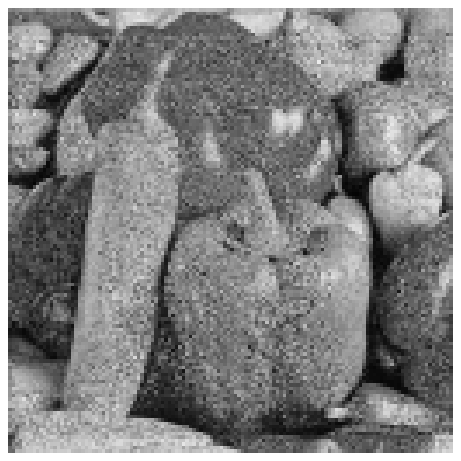}}
\\
\subfigure[]{\label{d-60-jpeg}\includegraphics[width=1.3in,height=1.3in]{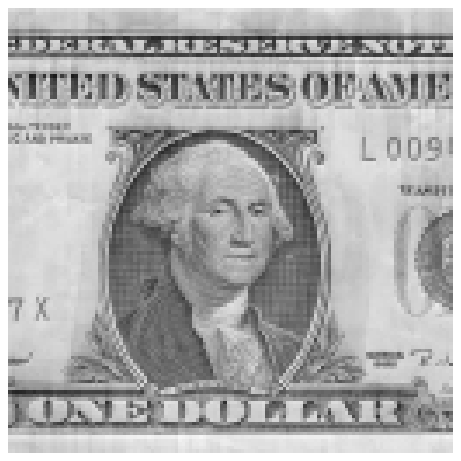}}
\hspace*{4mm}
\subfigure[]{\label{p-60-jpeg}\includegraphics[width=1.3in,height=1.3in]{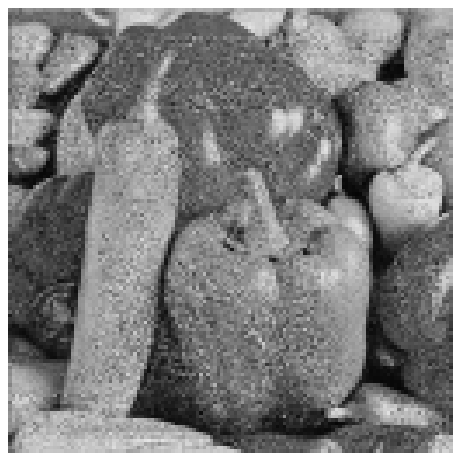}}
\\
\subfigure[]{\label{d-80-jpeg}\includegraphics[width=1.3in,height=1.3in]{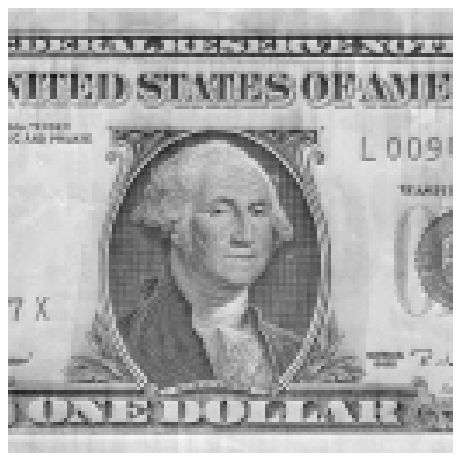}}
\hspace*{4mm}
\subfigure[]{\label{p-80-jpeg}\includegraphics[width=1.3in,height=1.3in]{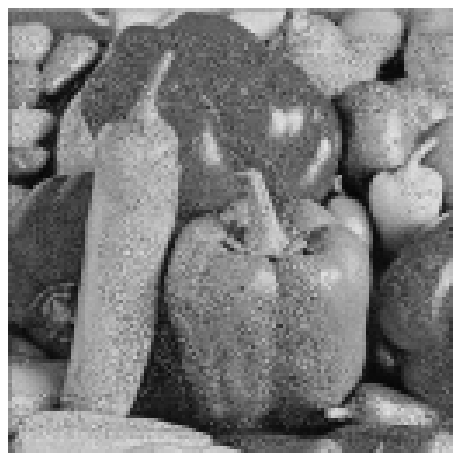}}
\caption{This panel shows the watermarked image on the left panel
and the extracted watermark on the right panel for
\subref{d-20-jpeg}, \subref{p-20-jpeg} JPEG compressed $20 \%$
watermarked image and Extracted watermark \subref{d-40-jpeg},
\subref{p-40-jpeg} JPEG compressed $40 \%$ watermarked image and
Extracted watermark \subref{d-60-jpeg}, \subref{p-60-jpeg} JPEG
compressed $60 \%$ watermarked image and Extracted watermark
\subref{d-80-jpeg}, \subref{p-80-jpeg} JPEG compressed $80 \%$
watermarked image and Extracted watermark.} \label{JPEG}
\end{center}
\end{figure}

JPEG compression \cite{jpegbook} is one of the popular standards for encoding
of digital images.
This is also an important test that any watermarking
algorithm has to defend itself against. We subject
Fig. \ref{d_EMB_0.0018} to various percentages of JPEG compression and
consequently extract the watermark from it. Figures \ref{d-20-jpeg},
\ref{d-40-jpeg}, \ref{d-60-jpeg}, and
\ref{d-80-jpeg} are JPEG compressed at $20\%$, $40\%$, $60\%$, and
$80\%$  respectively. The extracted watermarks from the aforementioned
compressed images are shown in Figs.
\ref{p-20-jpeg}, \ref{p-40-jpeg}, \ref{p-60-jpeg}, and
\ref{p-80-jpeg} respectively. From the PSNR values indicated in the table, it is
clear that the quality of the images extracted as well as the
embedded is better when the compression is less. This seems normal
since JPEG is a lossy compression method, the higher the
compression, the more is the loss in information of the image and
vice versa.

\begin{figure}
\begin{center}
\subfigure[]{\label{jp2one}\includegraphics[width=1.3in,height=1.3in]{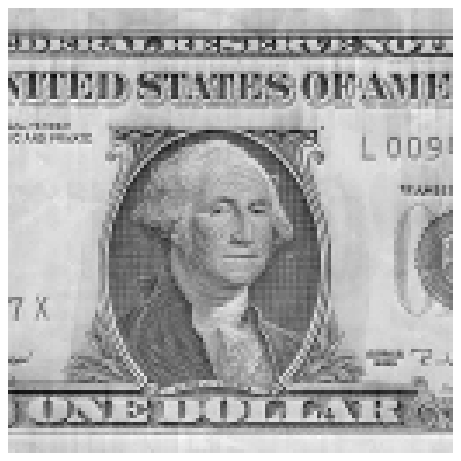}}
\hspace*{4mm}
\subfigure[]{\label{jp2two}\includegraphics[width=1.3in,height=1.3in]{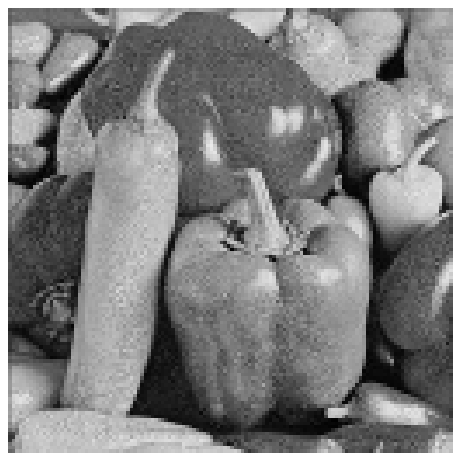}}
\caption{\subref{jp2one} Watermarked image after JPEG 2000
compression (Compression Ratio 2) and \subref{jp2two} Extracted
watermark.} \label{jp2}
\end{center}
\end{figure}

\begin{figure}
\begin{center}
\subfigure[]{\label{PSNRjp2one}\includegraphics[width=2.6in,height=2.0in]{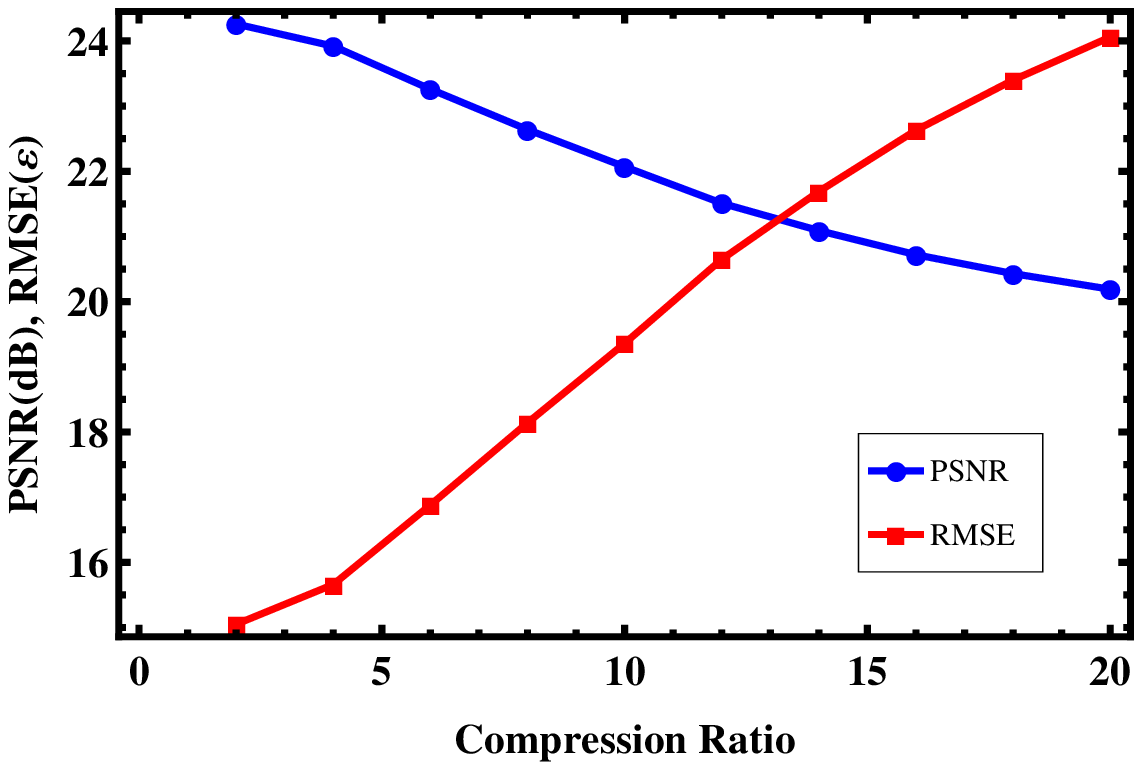}}
\hspace*{4mm}
\subfigure[]{\label{PSNRjp2two}\includegraphics[width=2.6in,height=2.0in]{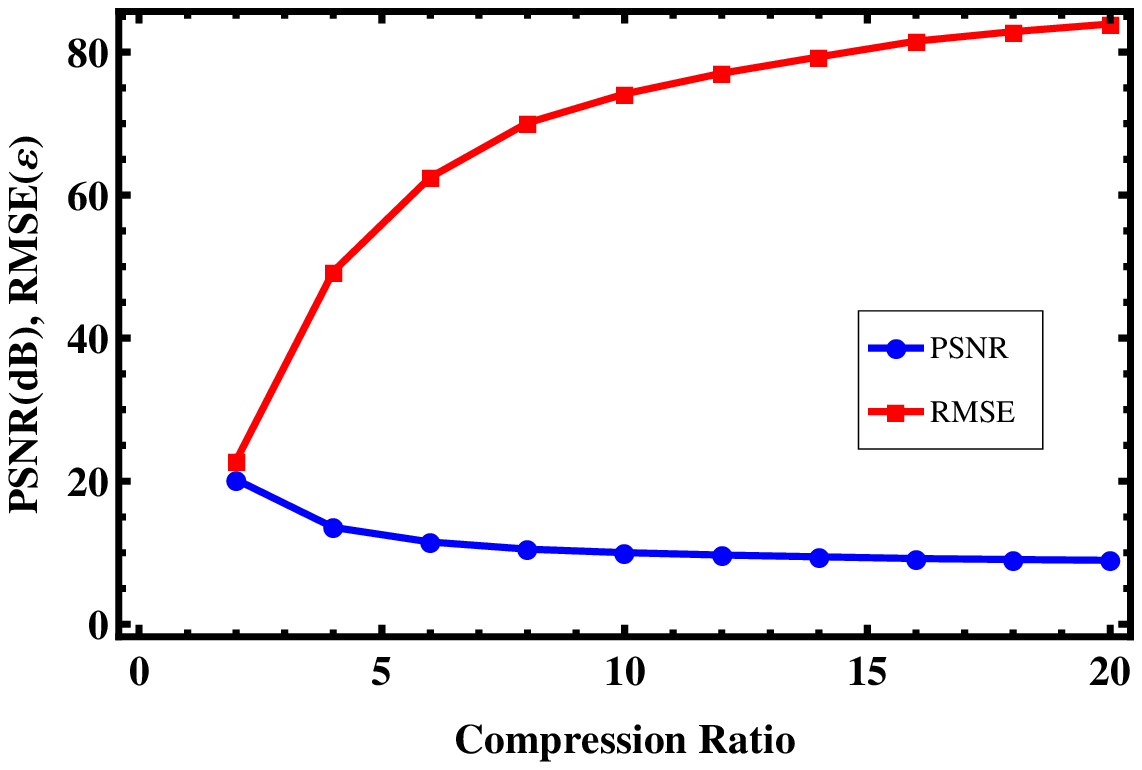}}
\caption{Graph depicting the variation of $p$ and $\e$, for various
compression ratios taken in the $x$ axis, \subref{PSNRjp2one}
Between the original and the watermarked images and
\subref{PSNRjp2two} Between the watermark and the extracted images.}
\label{psnr-jp2}
\end{center}
\end{figure}

Now we focus on JPEG 2000 standard \cite{jp2book} and show the results of attacks using JPEG 2000.
This is a more recent standard for image compression and is
based on wavelets and is much more efficient in reducing the size of images.
Figures \ref{jp2one} and \ref{jp2two} show, the compressed watermarked image and
the result of the extraction from it respectively at a JPEG 2000 ratio valued at
two. The PSNR and RMSE plot for the watermarked image for various JPEG 2000
compression ratios is given in Fig. \ref{PSNRjp2one}. Similarly the plot for the
extracted watermark is given in Fig. \ref{PSNRjp2two}.
Just like the JPEG case, the results show a deterioration in the
quality of the watermarked as well as of the extracted watermark images as the
compression ratio is increased. However, the watermark
can still be obtained even for very high compression ratios which could be
attributed to the robustness of the algorithm.

\subsection{Median Filter}

\begin{figure}
\begin{center}
\subfigure[]{\label{d-median}\includegraphics[width=1.3in,height=1.3in]{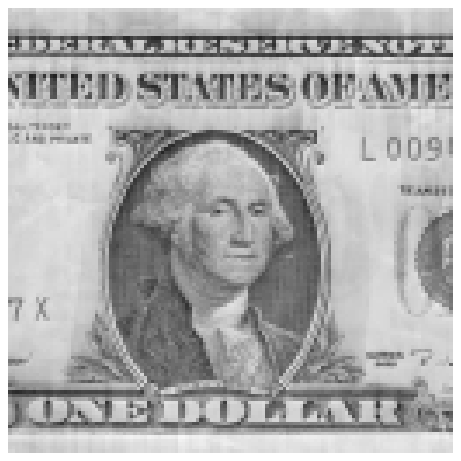}}
\hspace*{4mm}
\subfigure[]{\label{p-median}\includegraphics[width=1.3in,height=1.3in]{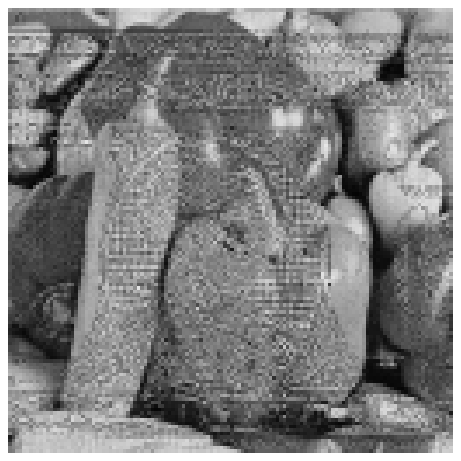}}
\caption{\subref{d-median} Median filtered watermarked image and
\subref{p-median} Extracted watermark.} \label{m-filter}
\end{center}
\end{figure}

Median filtering is used to remove outliers without reducing the sharpness of the
image. It is similar to an averaging filter in which the value of the output
pixel is the mean of the pixel values in the neighborhood of the corresponding
input pixel. However, in the present method of median filtering, as one might
have guessed, the value of an output pixel is determined by the median of the
neighborhood pixels. The extracted watermark, Fig. \ref{p-median}, from the
median filtered Fig. \ref{d-median} shows that the proposed scheme is robust to
such an attack.

\subsection{Rotation}

\begin{figure}
\begin{center}
\subfigure[]{\label{d-rotate}\includegraphics[width=1.3in,height=1.3in]{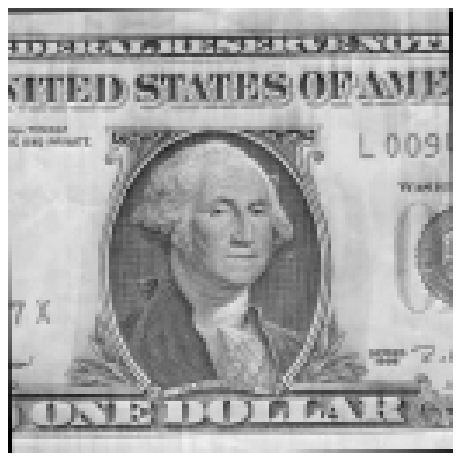}}
\hspace*{4mm}
\subfigure[]{\label{p-rotate}\includegraphics[width=1.3in,height=1.3in]{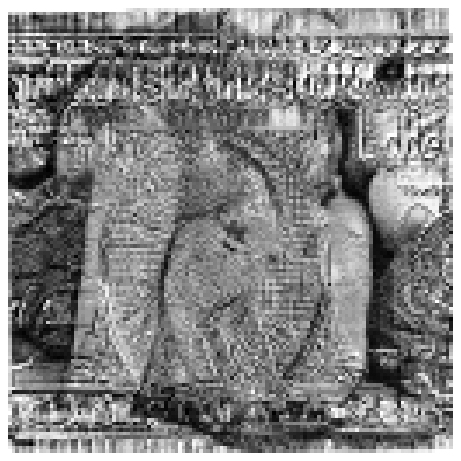}}
\caption{\subref{d-rotate} $1$ Degree rotated watermarked image and
\subref{p-rotate} Extracted watermark.} \label{rotate}
\end{center}
\end{figure}

Figure \ref{d-rotate} shows the result of rotating the watermarked image,
Fig. \ref{d_EMB_0.0018}, through one degree in the anti-clockwise direction. After this
has been done we crop the four corners of the rotated
image in order to keep the same size as the original image. Now when the extraction
algorithm is applied, the resulting extracted watermark is shown in Fig. \ref{p-rotate}.
Tests have been done for rotations by $2$ and $3$ degrees too. In these cases we are
able to retrieve more than $50\%$ of the watermark after extraction.

\subsection{Gamma Correction}

In general many output devices, say computer displays, have a nonlinear response
to that of the input signal. Gamma correction is applied to compensate for
this effect. Typically, the response follows a power law; $x^\gamma$. In the case of
images, a gamma correction to it has the effect of compressing and expanding the pixel
value. When $\gamma < 1$ it is called compression and for $\gamma > 1$ it is called
expansion. Since, $\gamma$ is related to the value of the pixel, which
in turn directly translates into visual quality of images, a gamma correction leads
to brighter or darker images depending on the value of $\gamma$.

\begin{figure}
\begin{center}
\subfigure[]{\label{d-gamm-corr-0.95}\includegraphics[width=1.3in,height=1.3in]{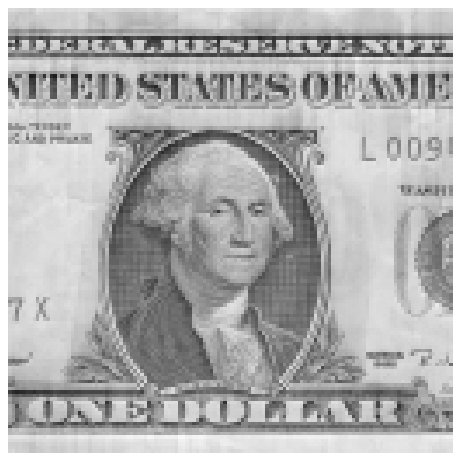}}
\hspace*{4mm}
\subfigure[]{\label{p-gamm-corr-0.95}\includegraphics[width=1.3in,height=1.3in]{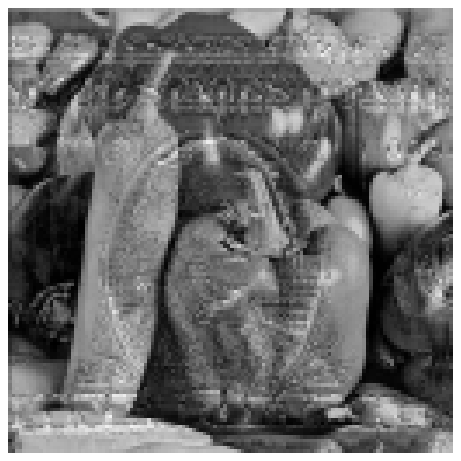}}
\\
\subfigure[]{\label{d-gamm-corr-1.09}\includegraphics[width=1.3in,height=1.3in]{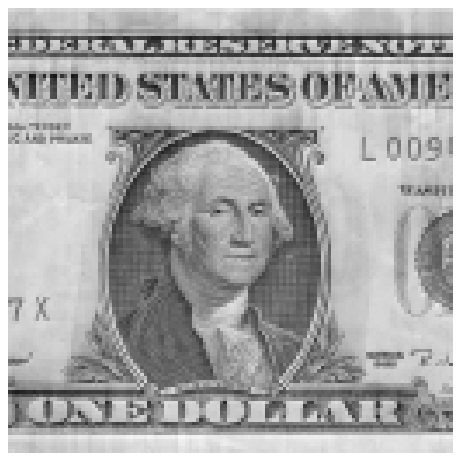}}
\hspace*{4mm}
\subfigure[]{\label{p-gamm-corr-1.09}\includegraphics[width=1.3in,height=1.3in]{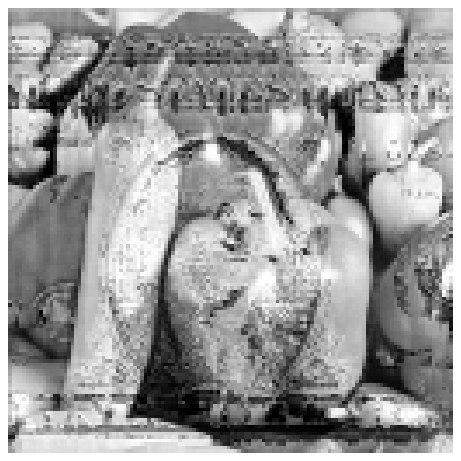}}
\caption{\subref{d-gamm-corr-0.95}, \subref{p-gamm-corr-0.95} Gamma
$0.95$ watermarked image and Extracted watermark
\subref{d-gamm-corr-1.09}, \subref{p-gamm-corr-1.09} Gamma $1.09$
watermarked image and Extracted watermark.} \label{g-corr}
\end{center}
\end{figure}

Figures \ref{d-gamm-corr-0.95} and \ref{p-gamm-corr-0.95} show the watermarked image
and the extracted watermark respectively after gamma compression with $\gamma=0.95$.
Figures \ref{d-gamm-corr-1.09} and \ref{p-gamm-corr-1.09} show the watermarked
and extracted images respectively after gamma expansion with $\gamma=1.09$.
In both the cases, the extracted watermark is clearly visible and PSNR value
is within the acceptable limits.

\subsection{Intensity Adjustment}

\begin{figure}
\begin{center}
\subfigure[]{\label{d-intensity}\includegraphics[width=1.3in,height=1.3in]{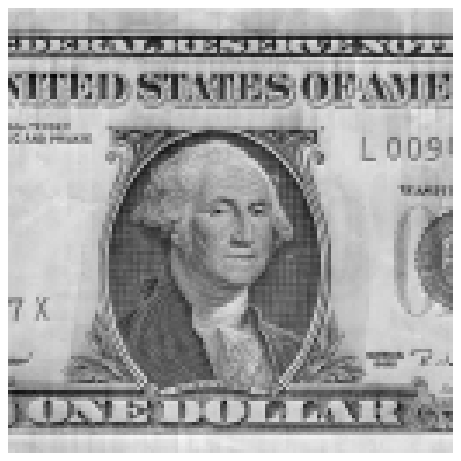}}
\hspace*{4mm}
\subfigure[]{\label{p-intensity}\includegraphics[width=1.3in,height=1.3in]{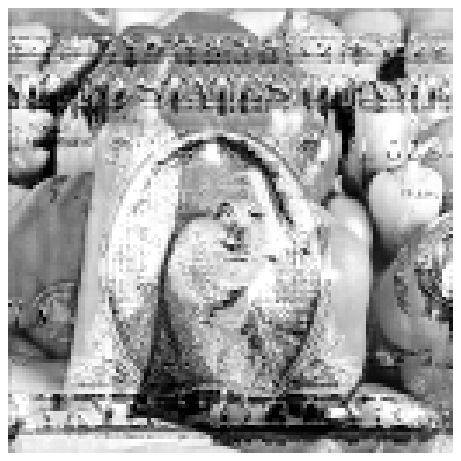}}
\caption{\subref{d-intensity} Intensity adjusted watermarked image
and \subref{p-intensity} Extracted watermark.} \label{i-adjust}
\end{center}
\end{figure}

Intensity adjustment is a technique of mapping an image's intensity values
to a new range. In the present example, we have mapped the watermarked image's
intensity to the range 0.1-1.0. As shown in Fig.
\ref{d-intensity} and \ref{p-intensity}, the watermark recovered
from the intensity adjusted image does not suffer much distortion.

\subsection{Gaussian Blur}

\begin{figure}
\begin{center}
\subfigure[]{\label{Gblurone}\includegraphics[width=1.3in,height=1.3in]{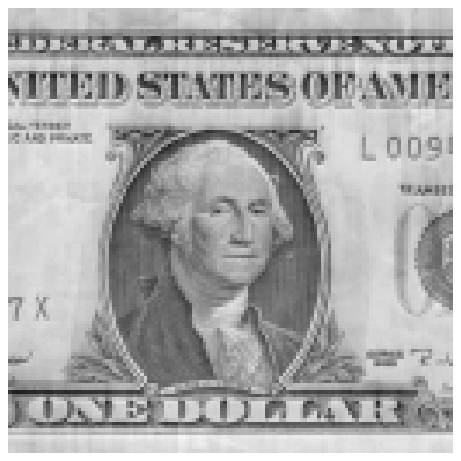}}
\hspace*{4mm}
\subfigure[]{\label{Gblurtwo}\includegraphics[width=1.3in,height=1.3in]{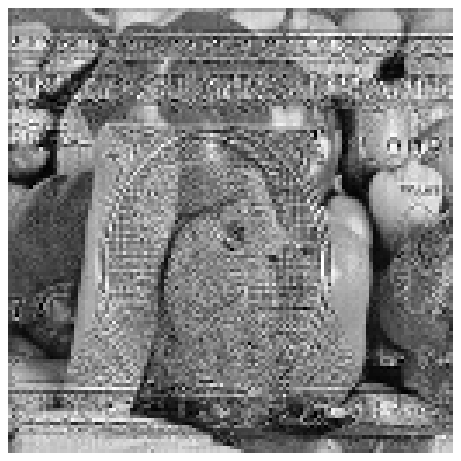}}
\caption{\subref{Gblurone} Watermarked image after Gaussian blur and
\subref{Gblurtwo} Extracted watermark.} \label{Gblur}
\end{center}
\end{figure}

The Gaussian blur filter, as implied by its very name, blurs objects. This filter
creates an output image after applying a Gaussian weighted average of the input pixels
around the location of each corresponding output pixel. Figure \ref{Gblurone} is the
Gaussian blurred watermarked image and Fig. \ref{Gblurtwo} is the watermark
extracted from it.

Blurring can also be achieved by a simple low-pass filter. Its mainly applied
to average
out the rapid changes in the intensity of the image. This is achieved by
calculating the average of a pixel and all of its eight immediate neighbors. This
average is then used instead of the original value of the pixel. This process is
repeated for every pixel of the image. Since, this is very similar to the Gaussian
blur attack, already presented, we will not give the resulting images of this experiment.
Suffice to say that the extraction of the watermark is possible.

\subsection{Contrast Enhancement}

Contrast enhancement is the process by which the contrast of an image is modified.
This can significantly improve the visual quality of the image. In this process
the darker colors are made more dark and the lighter regions more light. This is
achieved in the following way. We choose a particular cut off for the darker as well
as the lighter regions and all the pixel values that are smaller and larger than
this cut off are set to their mnimimum and maximum values respectively.
For example, in the case of a greyscale image all pixel values
that are smaller than the lower bound, chosen apriori, are made black. Correspondingly
all pixel values larger than the chosen upper bound are set to $255$, i.e., made white.
The rest of the values that lie in between
this minimum and maximum bound are spread linearly on a $0$ to $255$ scale.

\begin{figure}
\begin{center}
\subfigure[]{\label{cont-enh-one}\includegraphics[width=1.3in,height=1.3in]{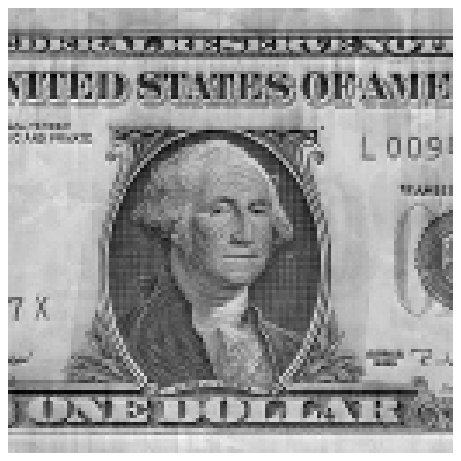}}
\hspace*{4mm}
\subfigure[]{\label{cont-enh-two}\includegraphics[width=1.3in,height=1.3in]{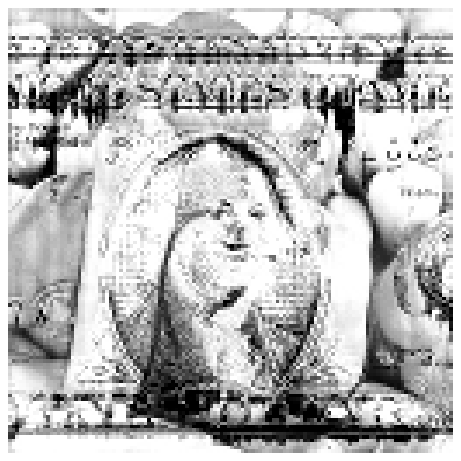}}
\caption{\subref{cont-enh-one} Watermarked image after applying
contrast enhancement and \subref{cont-enh-two} Extracted watermark.}
\label{cont-enh}
\end{center}
\end{figure}

The result of this attack on the watermarked image is shown in Fig.
\ref{cont-enh-one} and that of the extracted watermark is shown in
Fig. \ref{cont-enh-two}. The enhancement value chosen in the present
case is $0.75$. We have also checked it for another value $1.25$ and
from the PSNR and RMSE values given in the table, the quality of the
embedded and the extracted watermark actually improves with a higher
value of enhancement.

\subsection{Dilation}

\begin{figure}
\begin{center}
\subfigure[]{\label{dilate-one}\includegraphics[width=1.3in,height=1.3in]{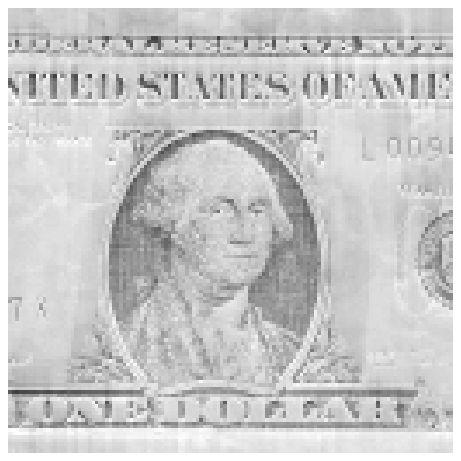}}
\hspace*{4mm}
\subfigure[]{\label{dilate-two}\includegraphics[width=1.3in,height=1.3in]{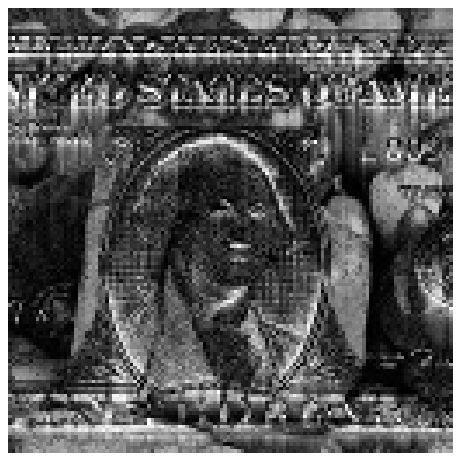}}
\caption{(a) Dilated watermarked image and (b) Extracted watermark.}
\label{dilate}
\end{center}
\end{figure}

Dilation in images can be thought of as a morphological operation. When it is
applied to grayscale images, the bright regions that are surrounded by dark regions
increase in size, whereas the dark regions having bright regions around it reduce
in size. The effect is very prominent at places, in the image, where the intensity
changes rapidly. Regions where the intensity is fairly uniform remain largely
unaffected except at the edges. Thus the overall effect of this operation is to
whiten the whole image. The results are shown in Figs. \ref{dilate-one} and \ref{dilate-two}.

\subsection{Scaling}

\begin{figure}
\begin{center}
\subfigure[]{\label{Uni-Rescale-one}\includegraphics[width=1.3in,height=1.3in]{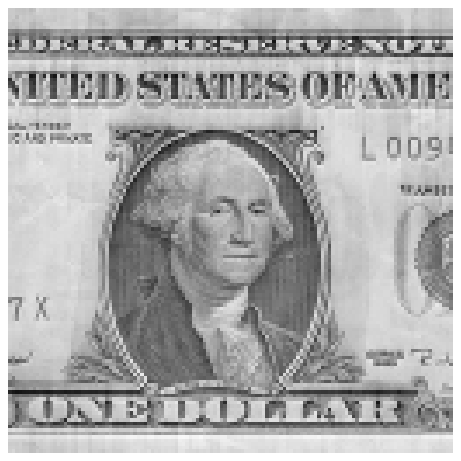}}
\hspace*{4mm}
\subfigure[]{\label{Uni-Rescale-two}\includegraphics[width=1.3in,height=1.3in]{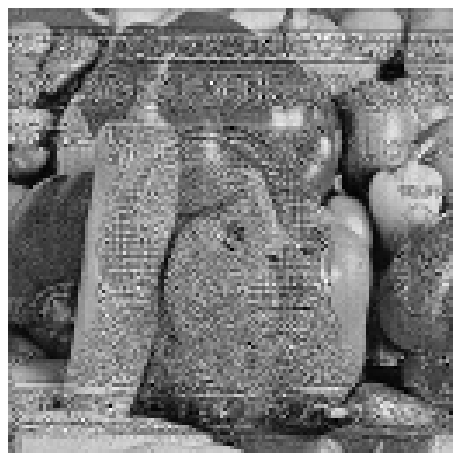}}
\\
\subfigure[]{\label{NU-Rescale-one}\includegraphics[width=1.3in,height=1.3in]{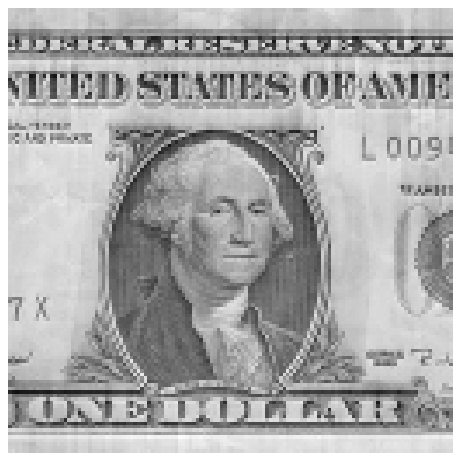}}
\hspace*{4mm}
\subfigure[]{\label{NU-Rescale-two}\includegraphics[width=1.3in,height=1.3in]{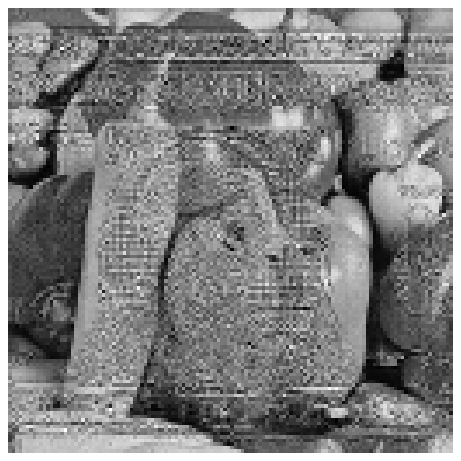}}
\caption{\subref{Uni-Rescale-one}, \subref{Uni-Rescale-two}
Uniformly rescaled watermarked image and Extracted watermark
\subref{NU-Rescale-one}, \subref{NU-Rescale-two} Non-uniformly
rescaled watermarked image and Extracted watermark.}
\label{rescaling}
\end{center}
\end{figure}

Scaling operation changes the size of images. This can be of two varieties,
namely, uniform and non-uniform. When the horizontal and vertical
directions are scaled by the same factor it is called uniform scaling.
In contrast to this, non-uniform scaling uses different scaling factors for the
horizontal and vertical directions. This latter type of scaling changes the aspect
ratio of the image unlike the former. The watermarked and extracted images after
uniform scaling are shown in Figs. \ref{Uni-Rescale-one} and
\ref{Uni-Rescale-two} respectively. The watermarked image, Fig. \ref{NU-Rescale-one},
and the extracted watermark, Fig. \ref{NU-Rescale-two}, after non-uniform scaling
show that our watermarking procedure performs quite well.

\subsection{Image Difference}

\begin{figure}
\begin{center}
\includegraphics[width=1.3in,height=1.3in]{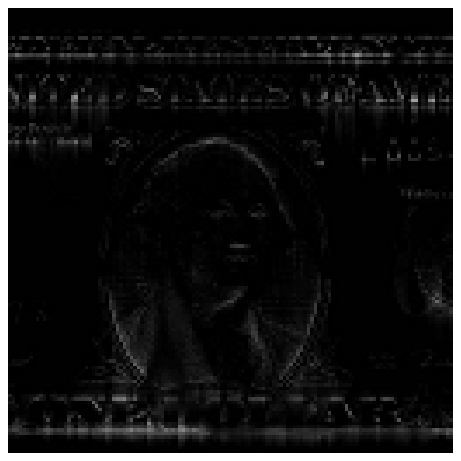}
\caption{Image difference between Fig. \ref{dollar512} and Fig.
\ref{d_EMB_0.0018}} \label{imagediff}
\end{center}
\end{figure}

The absolute difference between the original image and the
watermarked image, i.e, Fig. \ref{dollar512}  and Fig.
\ref{d_EMB_0.0018} is shown in Fig. \ref{imagediff} reveals that the
embedded watermark is completely invisible and we only see a
corrupted texture of the original image.
\begin{sidewaystable}
\begin{tabular}{|c|l|c|c|c|c|c|c|}
\hline \multirow{2}{*}{S.No.}  & \multirow{2}{*}{Type of Attack}  &
\multirow{2}{*}{Software Used} &
\multicolumn{2}{c|}{Dollar}  & \multicolumn{2}{c|}{Pepper}  \\
\cline{4-7}
               &            &               &      PSNR (dB)        &              RMSE ($\varepsilon$)     &    PSNR (dB)    &                RMSE ($\varepsilon$)          \\ \hline

   $1$.     &  Tune Sharpen (Directional) & Corel Graphics Suite 11 & 23.56 & 16.31 & 13.19 & 51.43  \\ \hline
   $2$.     &     Smoothing (50\%)& Corel Graphics Suite 11          & 22.53 & 18.36 & 10.88 & 67.14  \\ \hline
   $3$      &   Lowpass Blur (Radius 1)& Corel Graphics Suite 11       & 19.82 & 25.09 & 9.30 & 80.47  \\ \hline
   $4$      &  Gaussian Noise & MATLAB R2010a  & 22.32 & 18.81 & 9.83 & 75.71  \\ \hline
   $5$.     &  Gaussian Blur (Radius 1) & Corel Graphics Suite 11  & 19.18 & 27.02 & 8.95 & 83.85  \\ \hline
   $6$.     & Gamma correction $1.09$ & MATLAB R2010a  & 22.64 & 18.13 & 11.59 & 61.81  \\ \hline
   $7$.     & Gamma correction $0.95$ & MATLAB R2010a  & 24.92 & 13.94 & 15.17 & 40.95  \\ \hline
   $8$.     & Deinterlace (Even line, Interpolation) & Corel Graphics Suite 11  & 18.87 & 27.98 & 11.58 & 61.92  \\ \hline
   $9$.     & Contrast Enhancement 0.75 & Corel Graphics Suite 11  & 18.59 & 28.92 & 7.16 & 103.04  \\ \hline
   $10$.   & Contrast Enhancement 1.25 & Corel Graphics Suite 11  & 24.00 & 15.51 & 8.30 & 90.32  \\ \hline
   $11$.   & Dilation & GIMP 2.6.10  & 13.86 & 49.85 & 8.28 & 90.48  \\ \hline
   $12$.   & Erosion & GIMP 2.6.10  & 12.44 & 58.73 & 7.84 & 95.26  \\ \hline
   $13$.   & Local equalization (80, 80)& Corel Graphics Suite 11  & 11.90 & 62.47 & 6.38 & 112.61  \\ \hline
   $14$.   & Rotation 1 degree (Bilinear)& MATLAB R2010a  & 13.18 & 53.89 & 7.91 & 94.48  \\ \hline
   $15$.   & Additive Noise & MATLAB R2010a  & 18.56 & 29.01 & 13.37 & 50.59  \\ \hline
   $16$.   & Cropping & MATLAB R2010a  & 11.21 & 67.66 & 13.62 & 48.98  \\ \hline
   $17$.   & Median Filter (3$\times$3) & MATLAB R2010a  & 18.86 & 28.04 & 9.8 & 75.85  \\ \hline
   $18$.   & Intensity Adjustment & MATLAB R2010a  & 20.86 & 22.25 & 9.52 & 78.51  \\ \hline
   $19$.   & Uniform Rescaling (512-256-512) & Xn View  & 19.17 & 27.05 & 9.06 & 82.75  \\ \hline
   $20$.   & Non-uniform Rescaling (512-(320$\times$240)-512) & Xn View  & 5.78 & 126.37 & 9.10 & 82.38  \\ \hline
   $21$.   & JPEG $20\%$ & MATLAB R2010a  & 20.70 & 22.67 & 9.67 & 77.37  \\ \hline
   $22$.   & JPEG $40\%$ & MATLAB R2010a  & 21.98 & 19.57 & 10.47 & 70.85  \\ \hline
   $23$.   & JPEG $60\%$ & MATLAB R2010a  & 22.76 & 17.89 & 11.08 & 65.61  \\ \hline
   $24$.   & JPEG $80\%$ & MATLAB R2010a  & 23.63 & 16.18 & 12.66 & 54.68  \\ \hline
   $25$.   & JPEG2000 (Compression ratio $2$) & MATLAB R2010a  & 24.26 & 15.05 & 20.23 & 22.88  \\ \hline
   $26$.   & JPEG2000 (Compression ratio $4$) & MATLAB R2010a  & 23.92 & 15.66 & 13.55 & 49.33  \\ \hline
   $27$.   & JPEG2000 (Compression ratio $6$) & MATLAB R2010a  & 23.26 & 16.88 & 11.50 & 62.51  \\ \hline
   $28$.   & JPEG2000 (Compression ratio $8$) & MATLAB R2010a  & 22.64 & 18.14 & 10.50 & 70.11  \\ \hline
   $29$.   & JPEG2000 (Compression ratio $10$) & MATLAB R2010a  & 22.07 & 19.36 & 10.01 & 74.17  \\ \hline
   $30$.   & JPEG2000 (Compression ratio $12$) & MATLAB R2010a  & 21.51 & 20.65 & 9.68 & 77.06  \\ \hline
   $31$.   & JPEG2000 (Compression ratio $14$) & MATLAB R2010a  & 21.09 & 21.69 & 9.42 & 79.36  \\ \hline
   $32$.   & JPEG2000 (Compression ratio $16$) & MATLAB R2010a  & 20.72 & 22.63 & 9.19 & 81.56  \\ \hline
   $33$.   & JPEG2000 (Compression ratio $18$) & MATLAB R2010a  & 20.43 & 23.40 & 9.05 & 82.89  \\ \hline
   $34$.   & JPEG2000 (Compression ratio $20$) & MATLAB R2010a  & 20.19 & 24.06 & 8.93 & 83.96 \\
  \hline
\end{tabular}
\caption{Table showing different types of attacks performed on the
watermarked images. The PSNR and RMSE values of the watermarked as
well as the extracted image are given.}\label{Hilbert-tab}
\end{sidewaystable}

\section{Summary}

In this paper we have introduced a new watermarking scheme based
on the Hilbert transform of digital images. Due to inherent ambiguities
with the definitions of two-dimensional Hilbert transform, we apply
the one-dimensional Hilbert transforms column-wise to any image matrix
in which information is sought to be embedded. The Hilbert transformed
digital image is defined through its matrix of phase and amplitude values.
The main idea behind the proposed technique is that the
phase of the analytic signal associated with typical digital
images is generally a sparse matrix and hence offers large space for hiding
information in a highly imperceptible manner.

We have described embedding and extraction algorithms. The embedding is
performed in the Hilbert transformed domain. The Hilbert transform of
a watermark image is embedded in the Hilbert transformed domain of the
host image, after suitable scaling by a factor $\lambda$. An
efficient algorithm for computing the optimal scaling factor $\lambda$ has
been derived. The quality of the watermarked image and extracted
watermark have been shown to be good when measured in terms of PSNR and RMSE.
In addition, we have performed a large number of attacks to study the
robustness of our algorithm. The results are summarised in Table (\ref{Hilbert-tab}).
The proposed method is shown to be robust against additive noise,
cropping, gaussian noise, JPEG/JPEG 2000 compression, median
filtering, rotation, gamma correction, intensity adjustment,
gaussian blur, contrast enhancement, dilation, rescaling and image
difference. We would like to emphasise that calculating Hilbert transforms
can be efficiently done on personal computers too. It is not
computationally exacting. Thus, we believe
that our algorithm can also be modified to suit real-time applications
in an efficient manner.

\end{document}